\documentclass[aps,showpacs,superscriptaddress,preprint]{revtex4}%
\usepackage{mathrsfs}
\usepackage{amsfonts}
\usepackage{amsmath}
\usepackage{amssymb}
\usepackage{graphicx}%
\begin{document}
\title{Quasiparticle states and quantum interference induce by magnetic impurities on
a two-dimensional topological superconductor}
\author{Zhen-Guo Fu}
\affiliation{State Key Laboratory for Superlattices and Microstructures, Institute of
Semiconductors, Chinese Academy of Sciences, P. O. Box 912, Beijing 100083,
People's Republic of China}
\affiliation{LCP, Institute of Applied Physics and Computational Mathematics, P.O. Box
8009, Beijing 100088, People's Republic of China}
\author{Ping Zhang}
\thanks{zhang\_ping@iapcm.ac.cn}
\affiliation{LCP, Institute of Applied Physics and Computational Mathematics, P.O. Box
8009, Beijing 100088, People's Republic of China}
\author{Zhigang Wang}
\affiliation{LCP, Institute of Applied Physics and Computational Mathematics, P.O. Box
8009, Beijing 100088, People's Republic of China}
\author{Shu-Shen Li}
\affiliation{State Key Laboratory for Superlattices and Microstructures, Institute of
Semiconductors, Chinese Academy of Sciences, P. O. Box 912, Beijing 100083,
People's Republic of China}

\begin{abstract}
We theoretically study the effect of localized magnetic impurities on
two-dimensional topological superconductor (TSC). We show that the local
density of states (LDOS) can be tuned by the effective exchange field $m$, the
chemical potential $\mu$ of TSC, and the distance $\Delta r$ as well as
relative spin angle $\alpha$ between two impurities. The changes in $\Delta r$
between two impurities alter the interference and result in significant
modifications to the bonding and antibonding states. Furthermore, the
bound-state spin LDOS induced by single and double magnetic impurity
scattering, the quantum corrals, and the quantum mirages are also discussed.
Finally, we briefly compare the impurities in TSC with those in topological insulators.

\end{abstract}

\maketitle

\section{Introduction}

The study of topological insulators (TI) has attracted considerable
theoretical \cite{Kane, Bernevig,FuL,Moore, QiX,Zhanghj} and experimental
\cite{Konig,Hsieh1,Chen, Xia, Xue} interest over the past few years. These
studies have opened a door for exploring the promising potential applications
for spintronics with TI materials by their intrinsic strong spin-orbit
coupling (SOC) nature. Very recently, a series of superconductor (SC)/TI
hybrid structures, named as topological superconductors (TSCs) \cite{Fu, Qi},
were proposed to realize Majorana fermions by using the proximity effect of SC
and strong SOC of TI. Various transport proposals have been suggested to
detect and manipulate the Majorana fermions \cite{Fu1, Tanaka,
Lutchyn,Akhmerov, Fu2, Linder, Chung}. The experimental challenge along this
line remains to be overcome. In another way Shindou \textit{et al}.
\cite{Shindou} suggested that the surface-adsorbed magnetic impurities can be
effectively used to explore the Majorana fermions. Unfortunately, at present a
detailed and revealed study of the magnetic impurity effects on the surface
spectrum of TSC is still lacking, which is unexpected since a time reversal
breaking perturbation is the most natural way to reveal both the topological
and the superconducting properties of TSCs.

Motivated by the above-mentioned fact, in this paper we study the impurity
states and impurity interference on the two-dimensional (2D) TSC, which are
different from the ordinary SC systems because of the chirality and special
energy spectrum of the itinerant electron in TSC. We show that there exist
critical impurity-TSC spin coupling $m_{c}$ (following the appellation in Ref.
\cite{Salkola}) and critical chemical potential $\mu_{c}$ for a single
impurity, which correspond to unpaired zero-energy states. By analyzing the
spin local density of states (LDOS), a quantum transition from the
spin-unpolarized to the spin-polarized state is shown when the exchange
coupling parameter exceeds $m_{c}$. A clear understanding of quantum
interference is fundamental to our analysis of complex impurity structures.
Therefore, we illustrate that the distance and the relative spin angle between
two magnetic impurities can alter the quasiparticle interference effects,
resulting in significant modifications to the bonding and antibonding states.
For multiple impurities, we construct elliptical quantum corrals to study the
spectral properties of quantum mirages and the influence of quantum corrals on
the quantum interference effect between two magnetic impurities. We reveal
that the nonmagnetic (antiferromagnetic) corral has strong (weak) influence.
Finally, we briefly compare the impurity-TSC system with the impurity-TI
system, and we find that for the latter, the intragap quasiparticle states can
be arisen only when the scalar potential scattering are taken into account.
These findings, which can be detected by the scanning tunneling microscopy
(STM) and scanning tunneling spectroscopy (STS) techniques, may be useful for
determining the quasiparticle spectrum in TSC surface, and possess potential
applications in quantum computation.

\section{Theoretical formalism}

The 2D TSC by hybridizing an ordinary $s$-wave SC film with a TI film (say,
for example, Bi$_{2}$Se$_{3}$ film) is described by the Bogoliubov-de Gennes
Hamiltonian \cite{Fu}
\begin{equation}
H_{\text{BdG}}\left(  \boldsymbol{k}\right)  =\left(
\begin{array}
[c]{cc}%
h\left(  \boldsymbol{k}\right)  -\mu & i\Delta\sigma_{y}\\
-i\Delta^{\ast}\sigma_{y} & -h^{\ast}\left(  -\boldsymbol{k}\right)  +\mu
\end{array}
\right)  ,
\end{equation}
the basis for which is $\left(  c_{\boldsymbol{k}\uparrow},c_{\boldsymbol{k}%
\downarrow},c_{-\boldsymbol{k}\uparrow}^{\dagger},c_{-\boldsymbol{k}%
\downarrow}^{\dagger}\right)  ^{T}$. Here, for simplicity we just consider the
strong SOC term
\begin{equation}
h\left(  \boldsymbol{k}\right)  =v_{F}\left(  \sigma_{x}k_{x}+\sigma_{y}%
k_{y}\right)  .
\end{equation}
In Eq. (1) $\mu$ is the chemical potential, $\sigma_{x,y}$ are Pauli matrices
of electron spin, and the $s$-wave gap function is simply chosen to be a
constant, $\Delta$=$\Delta_{0}$. The impurity potential can be expressed as
\begin{equation}
V_{i}\mathtt{=}\frac{1}{2}(U_{i}\sigma_{0}\mathtt{+}J_{i}\boldsymbol{S}%
_{i}\mathtt{\cdot}\boldsymbol{\sigma})\tau_{z},
\end{equation}
where $\boldsymbol{S}_{i}\mathtt{=}S\boldsymbol{n}_{i}$ is the classical spin
(with its orientation vector $\boldsymbol{n}_{i}$) of the $i$th magnetic
impurity, $U_{i}$ and $J_{i}$ are the scalar potential and magnetic scattering
strengths, and the Pauli matrix $\tau_{z}$ acts on the particle-hole space. We
treat the impurity spin as a classical local effective exchange field
$m_{i}\mathtt{=}J_{i}S_{i}/2$ under mean-field approximation \cite{Liu}. By
employing the non-self-consistent $T\mathtt{-}$matrix method, we study the
quantum states and interferences induced by the magnetic impurities on the
TSC. The electronic Green's function in the presence of $N$ impurities is
written as
\begin{align}
G\left(  \boldsymbol{r},\boldsymbol{r}^{\prime},i\omega\right)   &
=G_{0}\left(  \boldsymbol{r},\boldsymbol{r}^{\prime},i\omega\right)
+\sum_{i,j=1}^{N}G_{0}\left(  \boldsymbol{r},\boldsymbol{r}_{i},i\omega
\right)  \nonumber\\
&  \times T\left(  \boldsymbol{r}_{i},\boldsymbol{r}_{j},i\omega\right)
G_{0}\left(  \boldsymbol{r}_{j},\boldsymbol{r}^{\prime},i\omega\right)
,\label{4}%
\end{align}
where the $T\mathtt{-}$matrix is given by the Bethe-Salpeter equation%
\begin{equation}
T\left(  \boldsymbol{r}_{i},\boldsymbol{r}_{j},i\omega\right)  =V_{i}%
\delta_{i,j}+V_{i}\sum_{l=1}^{N}G_{0}\left(  \boldsymbol{r}_{i},\boldsymbol{r}%
_{l},i\omega\right)  T\left(  \boldsymbol{r}_{l},\boldsymbol{r}_{j}%
,i\omega\right)  \label{5}%
\end{equation}
with the $4\mathtt{\times}4$ materix $G_{0}\left(  \boldsymbol{r}%
,\boldsymbol{r}^{\prime},i\omega\right)  $ obtained from the Fourier
transformation of the unperturbed Green's function
\begin{equation}
G_{0}\left(  \boldsymbol{k},i\omega\right)  =\left[  i\omega-H_{\text{BdG}%
}\right]  ^{-1}%
\end{equation}
for free TSC system. At half filling $\mu\mathtt{=}0$, one can find an
analytical expression of $G_{0}\left(  \mathbf{r}\mathtt{=}0,i\omega\right)  $
for $\left\vert \omega\right\vert \mathtt{<}\Delta_{0}$, written as
\begin{align}
G_{0}\left(  \mathbf{r}=0,i\omega\right)   &  =\frac{1}{4\pi\varrho v_{F}^{2}%
}\ln\left(  \frac{\Delta_{0}^{2}+\omega^{2}+v_{F}^{2}k_{c}^{2}}{\Delta_{0}%
^{2}+\omega^{2}}\right)  \nonumber\\
&  \times\left(  i\omega\tau_{0}+\Delta_{0}\tau_{2}\sigma_{2}\right)
,\label{g1}%
\end{align}
where $\varrho\mathtt{=}1/S_{\square}$ is the planar density with $S_{\square
}$ the area of the STC surface, and $k_{c}$ is high momentum cutoff. Finally,
we could get
\begin{equation}
G_{0}\left(  \mathbf{r}=0,\omega\right)  =-A\left(  \omega\right)  \left(
\omega\tau_{0}+\Delta_{0}\tau_{2}\sigma_{2}\right)  /D,
\end{equation}
where $A\left(  \omega\right)  \mathtt{=}\frac{2}{D}\ln\left(  D/\sqrt
{\Delta_{0}^{2}\mathtt{-}\omega^{2}}\right)  $ for $\Delta_{0}\mathtt{\ll}D$
with $D\mathtt{=}v_{F}k_{c}$. Here, we have used $4\pi\varrho v_{F}%
^{2}\mathtt{=}D^{2}$ with $\varrho\mathtt{=}k_{c}^{2}/4\pi$.

The total Green's function $G\left(  \boldsymbol{r},\boldsymbol{r}^{\prime
},\omega\right)  $ in Eq. (\ref{4}) can be rewritten as
\begin{equation}
G\left(  \boldsymbol{r},\boldsymbol{r}^{\prime},\omega\right)  =G_{0}\left(
\boldsymbol{r},\boldsymbol{r}^{\prime},\omega\right)  +\mathcal{G}%
_{0}\mathcal{TG}_{0}^{\prime}, \label{gt}%
\end{equation}
where the $4\mathtt{\times}4N$ matrix
\begin{equation}
\mathcal{G}_{0}=\left(
\begin{array}
[c]{cccc}%
G_{0}\left(  \boldsymbol{r},\boldsymbol{r}_{1},\omega\right)  & \cdots &
\cdots & G_{0}\left(  \boldsymbol{r},\boldsymbol{r}_{N},\omega\right)
\end{array}
\right)
\end{equation}
and the $4N\mathtt{\times}4$ matrix%
\begin{equation}
\mathcal{G}_{0}^{\prime}=\left(
\begin{array}
[c]{cccc}%
G_{0}\left(  \boldsymbol{r}_{1},\boldsymbol{r}^{\prime},\omega\right)  &
\cdots & \cdots & G_{0}\left(  \boldsymbol{r}_{N},\boldsymbol{r}^{\prime
},\omega\right)
\end{array}
\right)  ^{T},
\end{equation}
denote the propagation of electrons from the STM tip to the impurities as well
as from the impurities to the STM tip. Here, the $4N\mathtt{\times}4N$ matrix
$\mathcal{T}$ in Eq. (\ref{gt}) is a rewritten form of the Eq. (\ref{5}) for
the $T\mathtt{-}$matrix,%
\begin{equation}
\mathcal{T}=\mathcal{H}_{imp}+\mathcal{H}_{imp}\mathcal{G}_{0}\mathcal{T}%
\end{equation}
with the impurity Hamiltonian
\begin{equation}
\mathcal{H}_{imp}=\left(
\begin{array}
[c]{cccc}%
V_{1} & \mathbf{0} & \cdots & \mathbf{0}\\
\mathbf{0} & V_{2} & \cdots & \mathbf{0}\\
\vdots & \vdots & \ddots & \vdots\\
\mathbf{0} & \mathbf{0} & \cdots & V_{N}%
\end{array}
\right)
\end{equation}
containing all of the impurity scattering potentials $V_{i}$ ($i\mathtt{=}%
1,\cdots,N$). The information about the propagation between the impurities are
included in $\mathcal{G}_{0}$, which can be expressed as
\begin{equation}
\mathbf{G}_{0}=\left(
\begin{array}
[c]{cccc}%
G_{0}\left(  \boldsymbol{r}_{1},\boldsymbol{r}_{1}\right)  & G_{0}\left(
\boldsymbol{r}_{1},\boldsymbol{r}_{2}\right)  & \cdots & G_{0}\left(
\boldsymbol{r}_{1},\boldsymbol{r}_{N}\right) \\
G_{0}\left(  \boldsymbol{r}_{2},\boldsymbol{r}_{1}\right)  & G_{0}\left(
\boldsymbol{r}_{2},\boldsymbol{r}_{2}\right)  & \cdots & G_{0}\left(
\boldsymbol{r}_{2},\boldsymbol{r}_{N}\right) \\
\vdots & \vdots & \ddots & \vdots\\
G_{0}\left(  \boldsymbol{r}_{N},\boldsymbol{r}_{1}\right)  & G_{0}\left(
\boldsymbol{r}_{N},\boldsymbol{r}_{2}\right)  & \cdots & G_{0}\left(
\boldsymbol{r}_{N},\boldsymbol{r}_{N}\right)
\end{array}
\right)  .
\end{equation}
As a result, one can obtain
\begin{equation}
\mathcal{T}=\frac{\mathcal{H}_{imp}}{I-\mathcal{H}_{imp}\mathbf{G}_{0}},
\end{equation}
where $I$ is a $4N\mathtt{\times}4N$ unity matrix. Finally, by using these
equations, one can easily finish the numerical calculations with $N$
impurities, and then get the LDOS $\rho\left(  \mathbf{r},E\right)  $ of TSC
with impurities, which is given by
\begin{equation}
\rho\left(  \boldsymbol{r},\omega\right)  =-\frac{1}{\pi}\operatorname{Im}%
\sum_{i=1}^{2}G_{ii}\left(  \boldsymbol{r},\boldsymbol{r},\omega\right)
\equiv-\frac{1}{\pi}\operatorname{Im}\text{Tr}G^{(p)}\left(  \boldsymbol{r}%
,\boldsymbol{r},\omega\right)  ,
\end{equation}
where $G^{(p)}\left(  \boldsymbol{r},\boldsymbol{r}^{\prime},\omega\right)  $
is the particle Green function. Furthermore, the spin LDOS is written as
\begin{equation}
\boldsymbol{s}\left(  \boldsymbol{r}\right)  =-\frac{1}{2\pi}\operatorname{Im}%
\text{Tr}[G^{(p)}\left(  \boldsymbol{r},\boldsymbol{r},\omega\right)
\boldsymbol{\sigma}].
\end{equation}
For definiteness we set $k_{c}\mathtt{=}0.3\mathtt{\sim}0.4$ \AA $^{-1}$,
$v_{F}\mathtt{=}2.55$ eV$\cdot$\AA , and $\Delta_{0}\mathtt{=}1$ meV, which
give out a coherence length $\xi\mathtt{=}v_{F}/\Delta_{0}\mathtt{=}255$ nm.
The lattice constance is chosen as $a_{0}\mathtt{=}4.25$ \AA , and
$0^{+}\mathtt{=}0.02$ meV in following calculations. Here the chosen value of
$\Delta_{0}$ is consistent with recent theoretical prediction \cite{Fu} and
experimental implication \cite{Sasaki2011} on Cu$_{x}$Bi$_{2}$Se$_{3}$.

\section{Results and Discussions}

\subsection{Single magnetic impurity}

Similar to the ordinary $s$-wave SC systems, the purely scalar potential
scattering cannot form impurity bound state in the gap of TSC. Therefore, we
focus on the magnetic scattering effects in the following calculations.
Firstly, we consider a single magnetic impurity in STC.

We know that the $T\mathtt{-}$matrix possesses poles at resonance frequencies
$\omega_{res}^{\left(  1,2\right)  }$, which can reflect the presence of bound
states. The spectroscopic evidences for bound states are a pair of peaks in
the LDOS, which changes with $m$ as well as $\mu$ of the impurity-TSC system.
As $m$ increases, $\omega_{res}^{\left(  1,2\right)  }$ approach the chemical
potential, and at the critical coupling $m_{c}$, $\omega_{res}^{\left(
1,2\right)  }$ become to be degenerate at the zero energy. If $\left\vert
\omega\right\vert $ is not close to $\Delta_{0}$, one have a reasonable
approximation $A\left(  \omega\right)  \mathtt{\approx}A\left(  0\right)  $,
and then from the Green's function $G_{0}\left(  \boldsymbol{r}\mathtt{=}%
0,\omega\right)  $ in Eq. (\ref{g1}), we can easily obtain the $T\mathtt{-}%
$matrix at half filling $\mu\mathtt{=}0$, written as%
\begin{equation}
T\left(  \omega\right)  =\left[  V^{-1}-G_{0}\left(  \boldsymbol{r}%
=0,\omega\right)  \right]  ^{-1} \label{T}%
\end{equation}
for a single impurity case, which gives out critical coupling
\begin{equation}
m_{c}\left(  U\right)  \simeq\frac{1}{2}\sqrt{4D^{2}/\left(  \Delta
_{0}A\left(  0\right)  \right)  ^{2}+U^{2}} \label{MC}%
\end{equation}
and the bound-state energies%
\begin{equation}
\omega_{res}^{\left(  1,2\right)  }=-2c_{m}\pm\sqrt{c_{u}^{2}+\Delta_{0}^{2}}%
\end{equation}
for $m\mathtt{<}m_{c}\left(  U\right)  $, where $c_{m}\mathtt{=}2Dm/\left[
A\left(  0\right)  \left(  4m^{2}\mathtt{-}U^{2}\right)  \right]  $ and
$c_{u}\mathtt{=}2DU/\left[  A\left(  0\right)  \left(  4m^{2}\mathtt{-}%
U^{2}\right)  \right]  $. It is clear from Eq. (\ref{MC}) that taking into
account the scalar potential $U$ in the impurity potential can effectively
increase the critical coupling $m_{c}\left(  U\right)  $.

\begin{figure}[ptb]
\begin{center}
\includegraphics[width=0.7\linewidth]{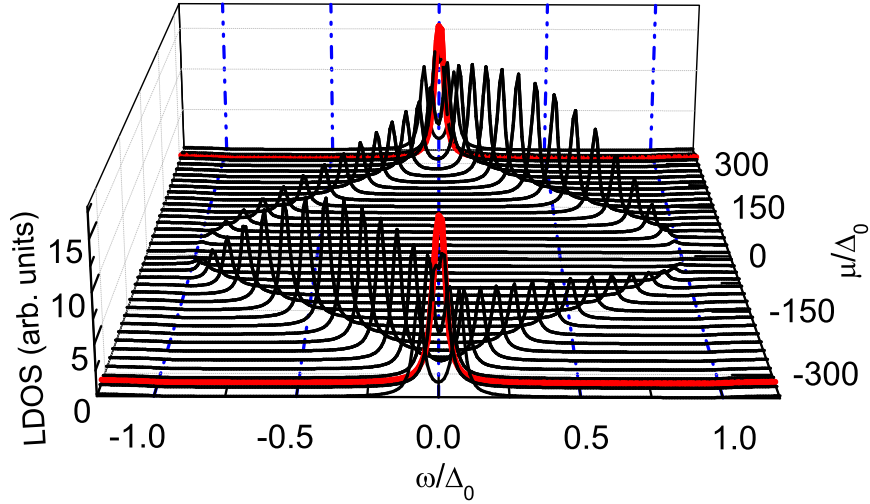}
\end{center}
\caption{ (Color online) The LDOS $\rho\left(  \boldsymbol{r}\mathtt{=}%
\mathbf{0},\omega\right)  $ of a single impurity located at origin with
$m\mathtt{=}0.8$ eV, $\boldsymbol{S}\Vert\hat{z}$, and $U\mathtt{=}0$, as a
function of frequency $\omega$ for a series of chemical potential $\mu$ in a
TSC. The unpaired zero-energy state are illustrated by red curves as $\mu$
increases to $\mu_{c}\mathtt{\approx}0.32$ eV.}%
\label{fig1}%
\end{figure}In the following calculations of $\mu\mathtt{\neq}0$, for
briefness we ignore the influence of the scalar potential term ($U$=$0).$ From
numerical calculations, we find that the critical magnetic moment
$m_{c}\mathtt{=}3.0,1.15$, $0.98$, and $0.71$ eV for $U\mathtt{=}0$
corresponding to the chemical potential $\mu\mathtt{=}0.05$, $0.2$, $0.25$,
and $0.37$ eV respectively. On the other hand, in Fig. \ref{fig1} we present
the LDOS at the impurity site $\boldsymbol{r}\mathtt{=}\left(  0,0\right)  $
as a function of $\mu$ for a fixed exchange field $m\mathtt{=}0.8$ eV. One can
see that the impurity resonance phenomenon approaches to vanish when tuning
the chemical potential to half filling ($\mu\mathtt{=}0$). The observable
intragap resonant states with energies of $\omega_{res}^{\left(  1,2\right)
}$ evolve from non-zero $\mu$, and their energy difference ($\left\vert
\omega_{res}^{\left(  2\right)  }\mathtt{-}\omega_{res}^{\left(  1\right)
}\right\vert $) decreases with increasing $\mu$. Particularly, at a critical
value $\mu_{c}$ of the chemical potential, the particle- and hole-like bound
states are degenerate at zero energy with equal spectral weight. In Fig.
\ref{fig1} we determine this critical chemical potential to be $\mu
_{c}\mathtt{=}\mathtt{\pm}320\Delta_{0}$ for the occurrence of zero-energy
bound states (see the red curves). Therefore, we can conclude that by tuning
the spin coupling or the host's chemical potential, the zero-energy bound
states could be detected on 2D TSC by employing spectral techniques such as STS.

\begin{figure}[ptb]
\begin{center}
\includegraphics[width=0.5\linewidth]{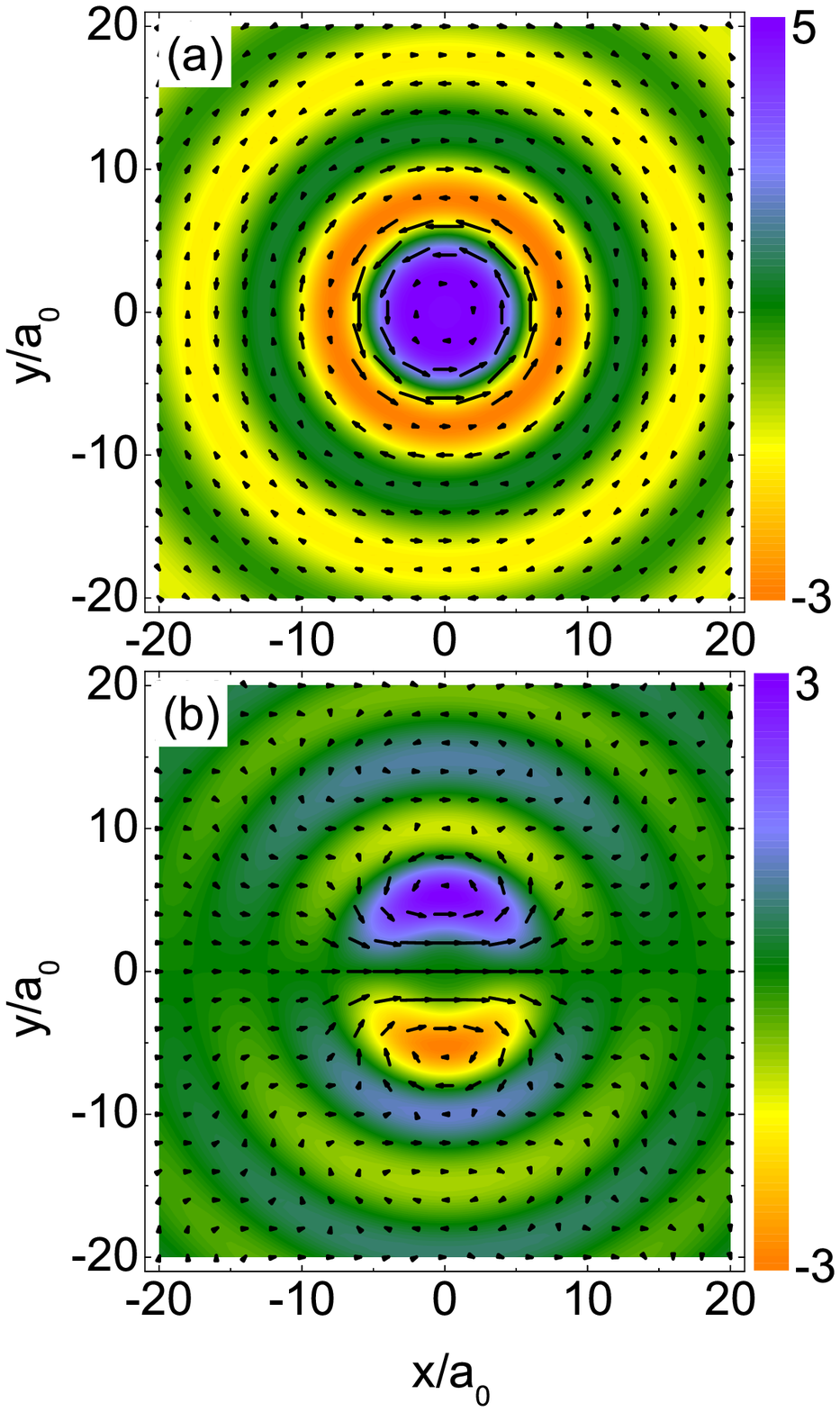}
\end{center}
\caption{ (Color online) The spatial distribution of spin LDOS at resonance
frequency $\omega_{res}\mathtt{=}-0.38\Delta_{0}$ for a single magnetic
impurity located at origin with $z$-direction (a) and $x$-direction
polarization. The background color denotes the $z$ component of spin LDOS
$s_{z}\left(  \boldsymbol{r},\omega\right)  $ while the arrow indicates the
$xy$-plane component $s_{\shortparallel}\left(  \boldsymbol{r},\omega\right)
$. The parameters are chosen as $m\mathtt{=}0.8$ eV and $\mu\mathtt{=}0.2$
eV.}%
\label{fig2}%
\end{figure}The occurrence of zero-energy bound states is a result of
competition among the pairing-condensation energy, the magnetic interaction,
and strong SOC. At these critical points, the ground state of STC may become
thermodynamically unstable, arising the quantum transformation of system from
spin-unpolarized to the spin-polarized state. As regarding to the spin
polarization, one should consider the spin LDOS around the impurity, from
which the spatially resolved spin polarization can be obtained by performing
integral over energy,
\begin{equation}
\left\langle \boldsymbol{s}\left(  \boldsymbol{r}\right)  \right\rangle
\mathtt{=}\mathtt{-}\frac{1}{\pi}\int_{-\infty}^{0}d\omega\boldsymbol{s}%
\left(  \boldsymbol{r},\omega\right)  ,
\end{equation}
as well as the total spin polarization $\left\langle \boldsymbol{s}%
\right\rangle \mathtt{=}\sum_{\boldsymbol{r}}\boldsymbol{s}\left(
\boldsymbol{r}\right)  $. The typical spatial distribution of spin LDOS is
plotted in Fig. \ref{fig2} at resonance frequency $\omega_{res}^{\left(
1\right)  }\mathtt{=}\mathtt{-}0.38\Delta_{0}$ for two choices of spin
orientations of the single magnetic impurity. Figure 2(a) is the case that the
local moment $\boldsymbol{S}\mathtt{\parallel}\hat{z}$, while Fig. 2(b)
corresponds to $\boldsymbol{S}\mathtt{\parallel}\hat{x}$. It is clear that the
impurity induces not only a $z$-direction spin polarization which decays with
oscillations, but also an $xy$-plane spin polarization in the present TSC
system. The spin LDOS possesses rotation symmetry about $z$-axis for a
$z$-direction impurity spin, shown as Fig. \ref{fig2}(a). However, the
rotation symmetry is broken when the impurity spin lies in plane. We find from
Fig. \ref{fig2}(b), for example, that the in-plane component
$s_{\shortparallel}\left(  \boldsymbol{r},\omega\right)  $ rotates
anticlockwise (clockwise) in the upper (lower) half-plane. These results can
be understood by the effective magnetic field $\boldsymbol{B}_{eff}%
\mathtt{=}v_{F}\boldsymbol{k}$ induced by the strong SOC in the 2D TSC, which
is similar to the case of a magnetic impurity on TI surface \cite{Liu}.
Because it is locked to the momentum of electron, with the moving of electron,
the spin undergoes a gyroscopic precession in the plane of perpendicular to
the propagation orientation, which arises corresponding slant of spin LDOS
$\boldsymbol{s}\left(  \boldsymbol{r},\omega\right)  $.

Taking a further step, corresponding to Fig. \ref{fig2}(a), we find that the
total spin polarization $\left\langle s_{z}\right\rangle \mathtt{\approx
}0.04\mathtt{<}1/2$, which indicates a negligibly small spin polarization.
This is consistent with our choice of $m$, which is lower than $m_{c}$; when
$m\mathtt{>}m_{c}$, an obvious spin polarization is observed. As a result, the
resonance peaks at negative (positive) energy region possesses a particle-like
and hole-like spin-chiral states $\left\vert p,\nearrow\right\rangle $ and
$\left\vert h,\searrow\right\rangle $, respectively, where $\nearrow$ and
$\searrow$ represent two different spin chiralities. This is different from
that on the ordinary $s$-wave SC surface with classical magnetic impurity,
where the particle/hole spin is parallel or antiparallel to the impurity spin
$\boldsymbol{S}$ \cite{Salkola,Morr1,Balatsky}. Notice that we have not found
a quantum transformation from spin-unpolarized to the spin-polarized state by
increasing $\mu$.

\subsection{Two magnetic impurities}

When the two impurities are close to each other, electrons will be scattered
by both impurities, resulting in quantum interference of electronic waves. The
interference effect is related to the scattering strength, the distance as
well as spin angle between two impurities. There are two important
consequences occurred due to the interference effect: (i) One is that the
interference can change the formation of the so-called bonding and antibonding
states $\left\vert p\left(  h\right)  \right\rangle _{b,a}\mathtt{=}\left(
\left\vert p\left(  h\right)  ,1\right\rangle \mathtt{\pm}\left\vert p\left(
h\right)  ,2\right\rangle \right)  /\sqrt{2}$ for the particle-like
(hole-like) states, with $\left\vert p,i\mathtt{=}1,2\right\rangle $ and
$\left\vert h,i\mathtt{=}1,2\right\rangle $ being the bound states of each
impurity; (ii) The other one is that the spin polarization will be altered
too, which can be observed from the spin LDOS patterns.

\begin{figure}[ptb]
\begin{center}
\includegraphics[width=0.7\linewidth]{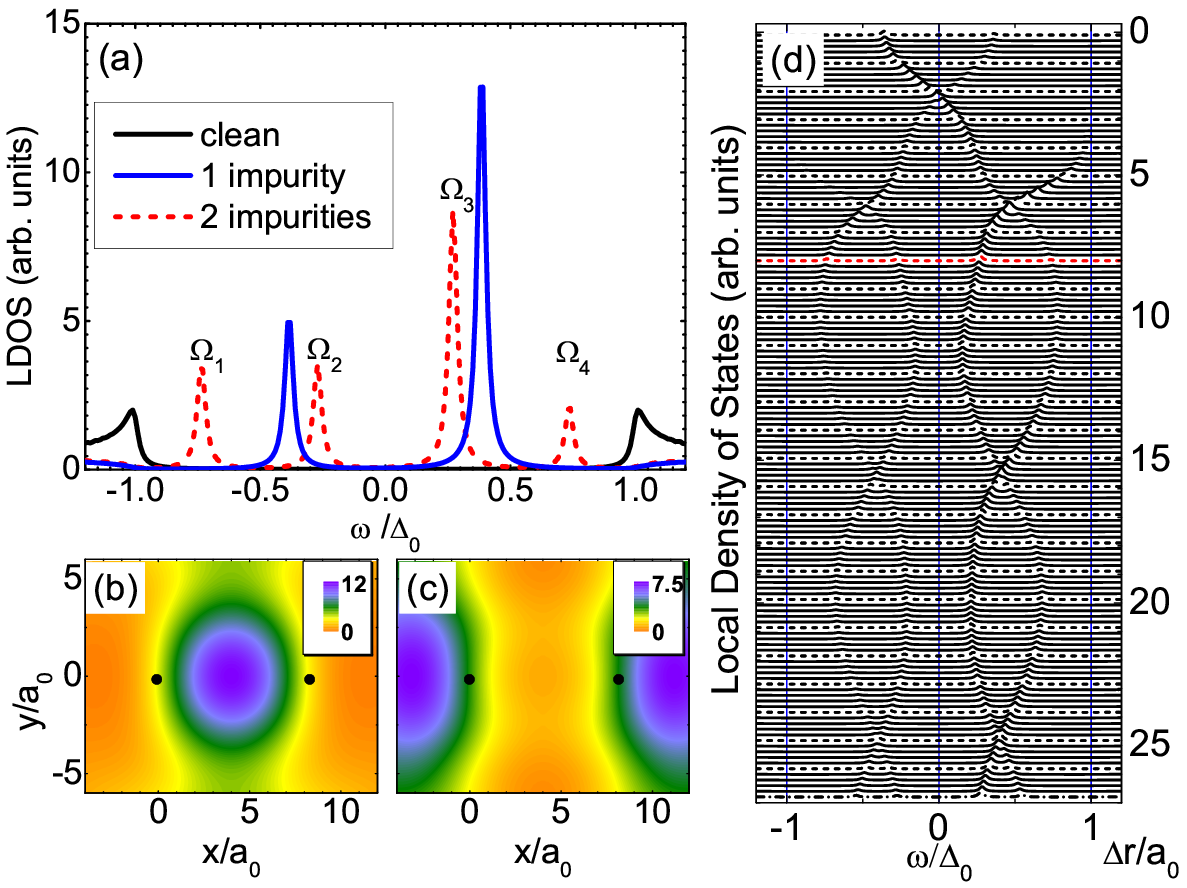}
\end{center}
\caption{ (Color online) (a) LDOS at impurity site for the clean surface
(black curve), a single impurity (blue curve), and two impurities (red dotted
curve) located at $\boldsymbol{r}_{1}\mathtt{=}\left(  0,0\right)  $ and
$\boldsymbol{r}_{2}\mathtt{=}\left(  8a_{0},0\right)  $ with parallel spins
along $z$-direction, respectively. (b) and (c) Spatial dependence of LDOS
corresponding to peaks $\Omega_{1}$ and $\Omega_{2}$ in (a), respectively. The
locations of impurities are denoted by black dots. (d) LDOS at $\boldsymbol{r}%
\mathtt{=}\mathbf{0}$ as a function of $\Delta r$ for two identical impurities
with parallel spins. The red dotted curve is for $\Delta r\mathtt{=}8a_{0}$. }%
\label{fig3}%
\end{figure}

Here, we focus on two identical magnetic impurities located at $\boldsymbol{r}%
_{1}\mathtt{=}(0,0)$ and $\boldsymbol{r}_{2}\mathtt{=}(\Delta r,0)$ with the
same exchange field $m\mathtt{=}0.8$ eV. Firstly, the LDOS for two impurities
located at $\boldsymbol{r}_{1}\mathtt{=}\left(  0,0\right)  $ and
$\boldsymbol{r}_{2}\mathtt{=}\left(  8a_{0},0\right)  $ is shown by the red
curve in Fig. \ref{fig3}(a), which exhibits four resonance peaks with peaks
$\Omega_{i=1,2}$ corresponding to the particle-like states $\left\vert
p\right\rangle _{b,a}$ and peaks $\Omega_{i=3,4}$ to the hole-like states
$\left\vert h\right\rangle _{b,a}$. If the interference of $\left\vert
p,1\right\rangle $ and $\left\vert p,2\right\rangle $ is constructive
(destructive) between two impurities, the bonding state $\left\vert
p/h\right\rangle _{b}$ ($\left\vert p/h\right\rangle _{a}$) is formed, which
can be observed from the spatial LDOS patterns. For this purpose, we plot in
Figs. \ref{fig3}(b) and \ref{fig3}(c) the spatial LDOS corresponding to the
particle-like state peaks $\Omega_{1}\mathtt{=}\mathtt{-}0.74\Delta_{0}$ and
$\Omega_{2}\mathtt{=}\mathtt{-}0.27\Delta_{0}$ respectively (the positions of
two impurities are indicated by black dots). It is obvious from Fig.
\ref{fig3}(b) that the LDOS are located in the middle region between two
impurities and reach maximum at $\Delta r\mathtt{=}\left(  4a_{0},0\right)  $,
therefore, peak $\Omega_{1}$ corresponds to a bounding state $\left\vert
p\right\rangle _{b}$. On the other hand, peak $\Omega_{2}$ should correspond
to an antibounding state $\left\vert p\right\rangle _{a}$ since the LDOS
vanishes at middle point between two impurities, see Fig. \ref{fig3}(c).

However, the interference pattern between the bonding and antibonding states
changes by varying the distance between the two impurities, i.e., the parity
and resonance energy of the two-impurity states oscillate with increasing
impurity separation. This can be observed from Fig. \ref{fig3}(d), which shows
the two-impurity distance $\Delta r$ dependency of LDOS. For instance,
$\left\vert p\right\rangle _{b}$ corresponds to peak $\Omega_{2}$ while
$\left\vert p\right\rangle _{a}$ corresponds to peak $\Omega_{1}$ when $\Delta
r\mathtt{=}5a_{0}$ or $20a_{0}$. The curves in Fig. \ref{fig3}(d) are moved in
order to illustrate the finer and more clear resonance frequencies. The
frequencies of the resonance oscillations, and at the same time the amplitude
of LDOS as well as their energy width changes. Moreover, we find that there
exists a critical distance $\Delta r_{c}\mathtt{\approx}2a_{0}$ similar to the
critical exchange parameter $m_{c}$ and critical chemical potential $\mu_{c}$
for single impurity case. At this point, the bound state energy of
particle-like $\left\vert p\right\rangle _{b}$ (hole-like $\left\vert
h\right\rangle _{b}$) crosses zero, and the state transforms into hole-like
$\left\vert h\right\rangle _{b}$ (particle-like $\left\vert p\right\rangle
_{b}$). Similar result has been theoretically observed in the ordinary
$s$-wave SC materials \cite{Morr1}. Whether or not a quantum spin polarization
transition occurs with increasing $\Delta r$ to exceed $\Delta r_{c}$ is not
clear, and further work are under way to study this interesting issue.
Besides, it is obvious that when the distance of two impurities is as small as
$\Delta r\mathtt{\lesssim}4a_{0}$, only two intragap bound states are found,
while there appears four non-degenerate bound states with increasing $\Delta
r$ to be larger than $4a_{0}$. With further increasing $\Delta r$, the
interference effect during the electron scattering processes by both
impurities gets so weak that the bound states become degenerate.
\begin{figure}[ptb]
\begin{center}
\includegraphics[width=0.7\linewidth]{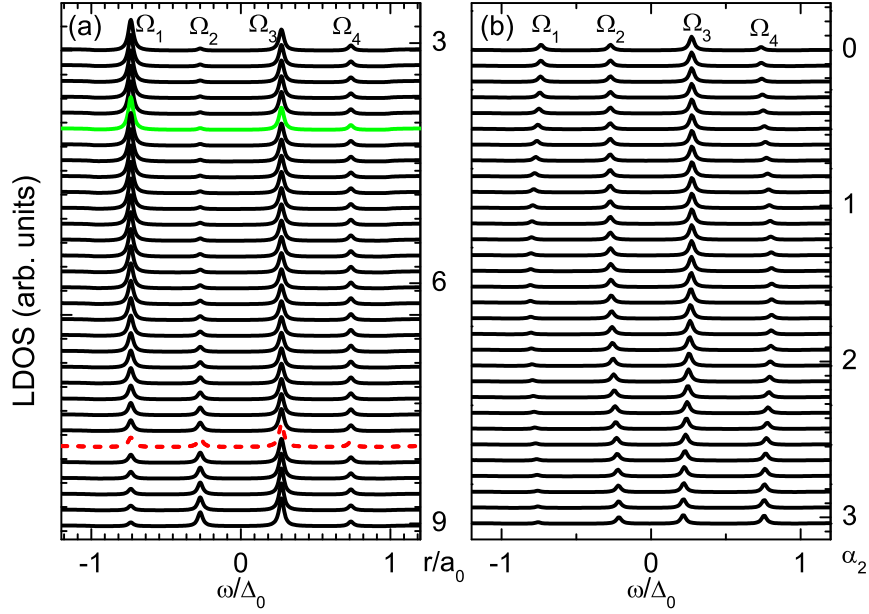}
\end{center}
\caption{ (Color online) (a) LDOS as a function of spatial position
$\boldsymbol{r}\mathtt{=}(r,0)$ for two impurities apart $8a_{0}$ with
parallel spins $S\hat{z}$. The red dashed curve represents the LDOS at
$\boldsymbol{r}_{2}$, and the green curve is for the midpoint $\boldsymbol{r}%
\mathtt{=}\left(  4a_{0},0\right)  $. (b) LDOS at $\boldsymbol{r}%
\mathtt{=}\mathbf{0}$ as a function of the angle $\alpha_{2}$ between two
impurity spins. The parameters are chosen as $m\mathtt{=}0.8$ eV and
$U\mathtt{=}0$.}%
\label{fig4}%
\end{figure}

An additional result of quantum interference between two impurities is the
change with the position of STM tip $\boldsymbol{r}$ in the number of
observable low-energy resonance peaks. For illustration we present in Fig.
\ref{fig4}(a) the LDOS as a function of $\boldsymbol{r}\mathtt{=}\left(
r,0\right)  $ for two impurities with parallel spins located at
$\boldsymbol{r}_{1}\mathtt{=}\left(  0,0\right)  $ and $\boldsymbol{r}%
_{2}\mathtt{=}\left(  8a_{0},0\right)  $. From four main low-energy intragap
resonances $\Omega_{i\mathtt{=}1,\mathtt{\cdots},4}$, one can find that the
resonance amplitude of particle-like (hole-like) peaks $\Omega_{1}$ and
$\Omega_{2}$ ($\Omega_{3}$ and $\Omega_{4}$) varies obviously (unobviously)
with $\boldsymbol{r}$. For example, for the uppermost curve $\boldsymbol{r}%
\mathtt{=}\left(  3a_{0},0\right)  $, the peak $\Omega_{1}$ ($\Omega_{2}$) is
sharp (smooth), while it is weaker (stronger) when the tip moves to
$\boldsymbol{r}\mathtt{=}\left(  9a_{0},0\right)  $, see the bottom curve in
Fig. \ref{fig4}(a). This is consistent with Fig. \ref{fig3}(b) because
$\Omega_{1}$ corresponds to a bonding state $\left\vert p\right\rangle _{b}$.
As mentioned above, the quantum interference effect could also be tuned
through varying the angle $\alpha_{2}$ between two impurity spins, as shown in
Fig. \ref{fig4}(b) for two impurities apart $8a_{0}$. We find that the
resonance frequencies $\Omega_{i}$ change with $\alpha_{2}$.
\begin{figure}[ptb]
\begin{center}
\includegraphics[width=0.5\linewidth]{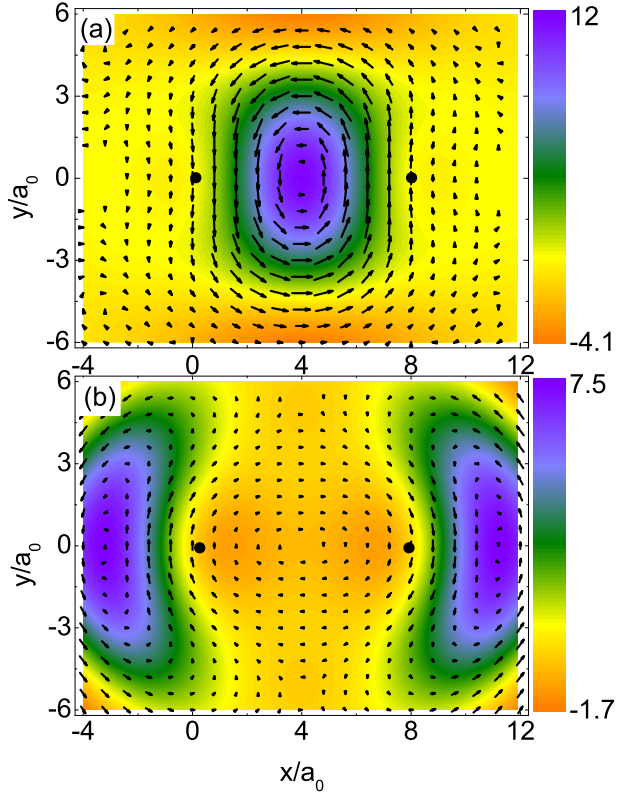}
\end{center}
\caption{ (Color online) Spin LDOS at resonance energy $\Omega_{1}$ (a) and
$\Omega_{2}$ (b) for two impurities. The parameters are the same as those in
Fig. \ref{fig3}.}%
\label{fig3added}%
\end{figure}

We also investigate the spin LDOS interference patterns of two impurities
apart $8a_{0}$ with parallel spins in the $z$-direction, and the results are
presented in Fig. \ref{fig3added}. It is found from Fig. \ref{fig3added}(a)
[\ref{fig3added}(b)] that on one hand, at the midpoint $\Delta r/2$ between
the two impurities, the $z$-component $s_{z}\left(  \boldsymbol{r}%
,\omega\right)  $ reaches its maximal (minimal) value, while the in-plane
component $s_{\shortparallel}\left(  \boldsymbol{r},\omega\right)  $ vanishes
when the bonding (antibonding) state is formed. Thus the spin in the bonding
(antibonding) particle-like state $\left\vert p\right\rangle _{b}$
($\left\vert p\right\rangle _{a}$) is completely parallel (antiparallel) at
$\Delta r/2$ to the impurity spins. On the other hand, the in-plane component
$s_{\shortparallel}\left(  \boldsymbol{r},\omega\right)  $ rotates
anticlockwise (clockwise) between the two impurities when the bonding
(antibonding) state is formed. These findings may contribute to potential
applications in the spin selection and also may be useful for analyzing the
bound-state electron mediated Ruderman-Kittel-Kasuya-Yosida (RKKY) spin-spin
interaction between magnetic impurities in TSC.

\subsection{Multiple impurities---quantum corral}

The nanostructrues, such as quantum corrals constructed by impurities on STC,
is also an issue of importance since different impurity structures usually
lead to different quantum interference behaviors. Here, as an illustrative
example, we briefly consider the spectral property of elliptical quantum
corrals with semimaxes $a\mathtt{=}10a_{0}$, $b\mathtt{=}8.66a_{0}$, and foci
$f_{\pm}\mathtt{=}\left(  \mathtt{\pm}5a_{0},0\right)  $, which are
constructed by $30$ nonmagnetic or magnetic impurities on STC. In the
following, we fix $\mu\mathtt{=}0.2$ eV and define $U_{i}\mathtt{=}0.8$ eV for
scalar impurities ($J_{i}\mathtt{=}0$), while $m_{i}\mathtt{=}0.8$ eV
($U_{i}\mathtt{=}0$) for magnetic ones. \begin{figure}[ptb]
\begin{center}
\includegraphics[width=0.7\linewidth]{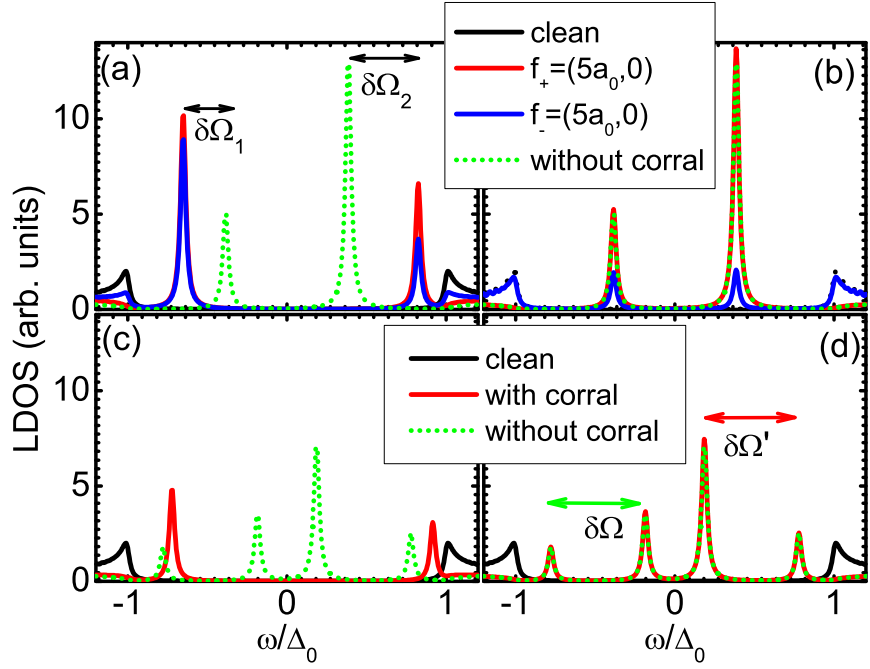}
\end{center}
\caption{ (Color online) (a-b): LDOS at the occupied focus $f_{+}$ (red curve)
and empty focus $f_{-}$ (blue curve) for a nonmagnetic $U\mathtt{=}0.8$ eV (a)
and antiferromagnetic $m\mathtt{=}0.8$ eV (b) corral with a magnetic impurity
located at $f_{+}$. (c-d): Splitting of the resonance peaks in LDOS at $f_{+}$
with (red curves) and without (green dotted curves) a nonmagnetic corral (c)
and a antiferromagnetic corral (d) respectively, for two magnetic impurities
with parallel spins. The chemical potential is $\mu=0.2$ eV.}%
\label{fig5}%
\end{figure}

Firstly, we consider the quantum mirage effect of a magnetic impurity that is
projected from the occupied into the empty focus of elliptical corrals. The
numerical spectral property of mirage versus the energy $\omega$ are presented
in Figs. \ref{fig5}(a) and \ref{fig5}(b) for nonmagnetic and antiferromagnetic
quantum corrals, respectively. The resonance peaks $\Omega_{1,2}$ of the
mirage at empty focus $f_{-}$ (blue curve) are in good agreement with those at
occupied focus $f_{+}$ (red curve), i.e., the mirage is clear. Comparing with
the case without corral, we find that the nonmagnetic corral unsymmetrically
shifts the particle-like and hole-like peaks. In Fig. \ref{fig5}(a) the
particle-like peak is shifted down by $\delta\Omega_{1}$=$0.27\Delta_{0}$,
while the hole-like peak is shifted up by $\delta\Omega_{2}$=$0.44\Delta_{0}$.
Surprisingly, although the mirage is obvious in antiferromagnetic corral [Fig.
\ref{fig5}(b)], the resonance peaks without corral are not shifted by
antiferromagnetic one in 2D TSC.

Furthermore, the influence of quantum corrals on the quantum interference
effects between two spin-parallel magnetic impurities apart $2\left\vert
f_{+}\right\vert $ on the TSC surface are also discussed. Although the
interference is dependent on many physical factors, such as spatial locations,
impurities scattering strength, relative angle between two impurity spins and
so on, for briefness we only consider the energy dependency of LDOS at $f_{+}$
of corrals. The results are shown in Figs. \ref{fig5}(c) and \ref{fig5}(d) for
nonmagnetic and antiferromagnetic corrals, respectively. One can find that the
influence of a nonmagnetic corral is prominent, see the red (green dotted)
curves in Figs. \ref{fig5}(c) which correspond to the presence (absence) of
corrals. However, the antiferromagnetic corral represents weak influence (the
splitting of bound state peaks is almost the same as that without corral, see
Fig. \ref{fig5}(d)). This result is different from that in ordinary SC system
\cite{Morr1}. We also find that the influence of corrals can be changed by
tuning the host chemical potential. When $\mu\mathtt{=}0.25$ eV, for example,
in the absence of corral the splitting of peaks is $\delta\Omega/\Delta
_{0}\mathtt{=}0.595$, while it increases to $\delta\Omega^{\prime}/\Delta
_{0}\mathtt{=}0.65$ in the presence of an antiferromagnetic corral.

\section{Comparison with impurity-TI system}

Before ending this paper, we would like to qualitatively compare the magnetic
impurities on TSC with those on TI surface. For briefness, we just consider
one- and two-impurity cases. In the presence of magnetic impurities, the TI
surface is described by%
\begin{align}
H  &  =h\left(  \boldsymbol{k}\right)  +\mu\sigma_{0}+\Delta_{0}\sigma
_{z}\nonumber\\
&  +\sum_{i=1}^{N}\left[  U\sigma_{0}+J_{i}S/2\sigma_{z}\right]  \delta\left(
\boldsymbol{r}-\boldsymbol{r}_{i}\right)  , \label{TI1}%
\end{align}
where $2\Delta_{0}$ in the third term is the energy gap induced by the
effective mass of Dirac fermions. The eigenvalues of the free part of Eq.
(\ref{TI1}) are
\begin{equation}
\epsilon_{k}=\pm\sqrt{\left(  v_{F}k\right)  ^{2}+\Delta_{0}^{2}}+\mu,
\end{equation}
which indicates that the chemical potential is not important for impurity
scattering in TI since it just move the levels up and down.
\begin{figure}[ptb]
\begin{center}
\includegraphics[width=0.7\linewidth]{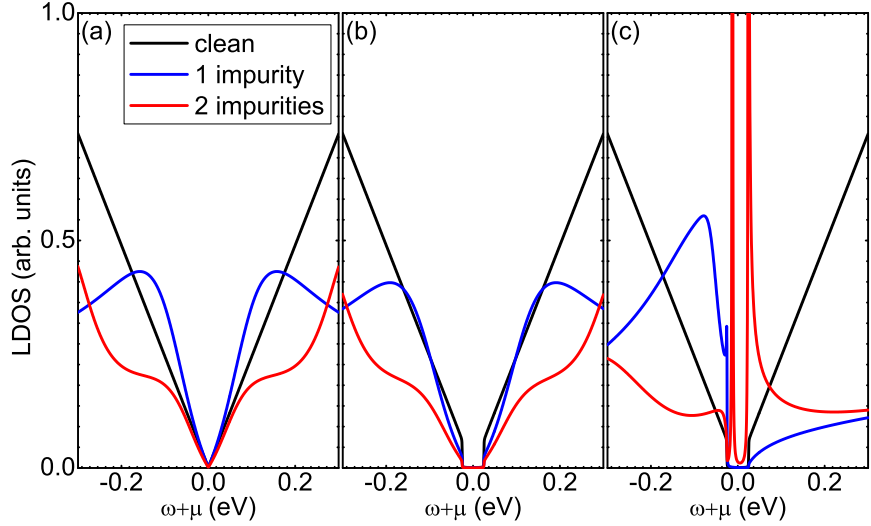}
\end{center}
\caption{ (Color online) (a) LDOS at impurity site $\boldsymbol{r}%
_{1}\mathtt{=}(0,0)$ on gapless (a) and gapped $\Delta_{0}\mathtt{=}25$ meV
(b) and (c) topological insulator surface. The black, blue, and red curves
correspond to clean surface, one single impurity, and two impurities apart
$2a_{0}$, respectively. In (a) and (b) $U\mathtt{=}0$, while in (c)
$U\mathtt{=}2$ eV. The magnetic moment is chosen as $m\mathtt{=}0.8$ eV.}%
\label{fig6}%
\end{figure}

The numerically calculated LDOS at impurity site $\boldsymbol{r}_{1}%
\mathtt{=}\left(  0,0\right)  $ for TI surface with $\mu\mathtt{=}0.2$ eV are
shown in Fig. \ref{fig6}, in which the black curves are for clean surface, and
the blue (red) curves are for TI surface with one (two) impurity. In Figs.
\ref{fig6}(a) and \ref{fig6}(b), we set the scalar scattering potential
$U$\texttt{=}$0$, while $U\mathtt{\neq}0$ in Fig. \ref{fig6}(c). One can see
that the low-energy LDOS is reshaped prominently by the impurity scattering on
massless as well as massive TI surface. For example, in Fig. \ref{fig6}(a) the
resonance peaks (blue curve) induced by one impurity are smoothed (blue curve)
due to the interference between two impurities. Similar to the impurity on STC
surface, the particle-hole symmetry holds on if we ignore $U$; Differing from
the case of TSC surface, however, we find that the magnetic impurity on TI
surface does not introduce intragap resonance peaks if one ignores $U$, see
Fig. \ref{fig6}(b). When the scalar potential is taken into account, we find
that, on one hand, the scalar potential introduces intragap resonance peaks,
which differs from the TSC case. On the other hand, the particle-hole symmetry
is now broken, as shown in Fig. \ref{fig6}(c). These resonance states could be
resolved from STM measurements, and thereby we hope the different impurity
effects in TSC and TI materials could be found in experiment.

\section{Conclusion}

In summary, we have studied the spectral properties of quasiparticle states
induced by localized classical magnetic impurities on 2D TSC. For the single
impurity case, the critical exchange field $m_{c}$ and critical chemical
potential $\mu_{c}$ are determined, and moreover, the spin LDOS is discussed.
Spin polarization transformation has been found in TSC system when
$m\mathtt{>}m_{c}$. For two-impurity scattering, we have discussed the
interference effects by changing the distance and relative spin angle between
two impurities. We have showed that the bound states can be changed by varying
the distance between the two impurities. The quantum mirages and the influence
of quantum corrals on the quantum interference effect between two impurities
have also been analyzed. The nonmagnetic corral exhibits stronger influence
than the antiferromagnetic one. Finally, we briefly compare the impurities in
TSC with those in TI. The results illustrated that the intragap resonance
states could be found in TI so long as the scalar scattering potential is
taken into account. These predictions, which could be observed by STM and STS
techniques, may be useful for exploring the electronic structures of TSC as
well as TI materials.

This work was supported by NSFC under Grants No. 90921003, No. 60776063, No.
60821061, and 60776061, and by the National Basic Research Program of China
(973 Program) under Grants No. 2009CB929103 and No. G2009CB929300.


\begin{thebibliography}{99}                                                                                               %


\bibitem {Kane}C. L. Kane and E. J. Mele, Phys. Rev. Lett. \textbf{95}, 146802 (2005).

\bibitem {Bernevig}B. A. Bernevig, T. L. Hughes, and S.-C. Zhang, Science
\textbf{314}, 1757 (2006).

\bibitem {FuL}L. Fu, C. L. Kane, and E. J. Mele, Phys. Rev. Lett \textbf{98},
106803 (2007).

\bibitem {Moore}J. E. Moore and L. Balents, Phys. Rev. B \textbf{75},
121306(R) (2007).

\bibitem {QiX}X.-L. Qi, T. L. Hughes, and S.-C. Zhang, Phys. Rev. B
\textbf{78}, 195424 (2008).

\bibitem {Zhanghj}H. Zhang, C.-X. Liu, X.-L. Qi, X. Dai, Z. Fang and S.-C.
Zhang, Nat. Phys. \textbf{5}, 438 (2009).

\bibitem {Konig}M. K\"{o}nig, S. Wiedmann, C. Br\"{u}ne, A. Roth, H. Buhmann,
L. W. Molenkamp, X.-L. Qi, S.-C. Zhang, Science \textbf{318}, 766 (2007).

\bibitem {Hsieh1}D. Hsieh, D. Qian, L. Wray, Y. Xia, Y. S. Hor, R. J. Cava and
M. Z. Hasan, Nature \textbf{452}, 970 (2008); D. Hsieh, Y. Xia, D. Qian, L.
Wray, F. Meier, J. H. Dil, J. Osterwalder, L. Patthey, A. V. Fedorov, H. Lin,
A. Bansill, D. Grauer, Y. S. Hor, R. J. Cava, and M. Z. Hasan, Phys. Rev.
Lett. \textbf{103}, 146401 (2009).

\bibitem {Chen}Y.-L. Chen, J. G. Analytis, J.-H. Chu, Z.-K. Liu, S.-K. Mo,
X.-L. Qi, H.-J. Zhang, D.-H. Lu, X. Dai, Z. Fang, S.-C. Zhang, I. R. Fisher,
Z. Hussain, Z.-X. Shen, Science \textbf{325}, 178 (2009).

\bibitem {Xia}Y. Xia, D. Qian, D. Hsieh, L. Wray, A. Pal, H. Lin, A. Bansil,
D. Grauer, Y. S. Hor, R. J. Cava, and M. Z. Hasan, Nat. Phys. \textbf{5}, 398 (2009).

\bibitem {Xue}P. Cheng, C. Song, T. Zhang, Y. Zhang, Y. Wang, J.-F. Jia, J.
Wang, Y. Wang, B.-F. Zhu, X. Chen, X. Ma, K. He, L. Wang, X. Dai, Z. Fang, X.
C. Xie, X. L. Q, C. X. Liu, S. C. Zhang, and Q. K. Xue, Phys. Rev. Lett.
\textbf{105}, 076801 (2010).

\bibitem {Fu}L. Fu and C. L. Kane, Phys. Rev. Lett. \textbf{100}, 096407 (2008).

\bibitem {Qi}X.-L. Qi, T. L. Hughes, and S.-C. Zhang, Phys. Rev. B
\textbf{82}, 184516 (2010).

\bibitem {Fu1}L. Fu and C. L. kane, Phys. Rev. B \textbf{79}, 161408 (2009).

\bibitem {Tanaka}Y. Tanaka, T. Yokoyama, and N. nagaosa, Phys. Rev. Lett.
\textbf{103}, 107002 (2009).

\bibitem {Lutchyn}R. Lutchyn, J. Sau, and S. Das Sarma, Phys. Rev. lett.
\textbf{105}, 077001 (2010).

\bibitem {Akhmerov}A. R. Akhmerov, J. Nilsson, and C. W. J. Beenakker, Phys.
Rev. Leet. \textbf{102} 9216404 (2009).

\bibitem {Fu2}L. Fu and C. L. Kane, Phys. Rev. lett. \textbf{102}, 216403 (2009).

\bibitem {Linder}J. Linder, Y. Tanaka, T. Yokoyama, A. Sudb{\o }, and N.
nagaosa, Phys. Rev. Lett. \textbf{104}, 067001 (2010).

\bibitem {Chung}S. B. Chung, X.-L. Qi, J. Maciejko, and S.-C. Zhang, Phys.
Rev. B \textbf{83}, 100512(R) (2011).

\bibitem {Shindou}R. Shindou, A. Furusaki, and N. Nagaosa, Phys. Rev. B
\textbf{82}, 180505(R) (2010).

\bibitem {Salkola}M. I. Salkola, A. V. Balatsky, and J. R. Schrieffer, Phys.
Re.v B \textbf{55}, 12648 (1997).

\bibitem {Liu}Q. Liu, C.-X. Liu, C. Xu, X.-L. Qi, and S.-C. Zhang, Phys. Rev.
Lett. \textbf{102}, 156603 (2009).

\bibitem {Sasaki2011}S. Sasaki, M. Kriener, K. Segawa, K. Yada, Y. Tanaka, M.
Sato, and Y. Ando, arXiv:1108.1101v1 (2011).

\bibitem {Morr1}D. K. Morr and N. A. Stavropoulos, Phys. Rev. B \textbf{67},
020502(R) (2003); Phys. Rev. Lett. \textbf{92}, 107006 (2004); N. A.
Stavropoulos and D. K. Morr, Phys. Rev. B \textbf{71}, 14050(R) (2005).

\bibitem {Balatsky}A. V. Balatsky, I. Vekhter, J.-X. Zhu, Rev. Mod. Phys.
\textbf{78}, 373 (2006).
\end{thebibliography}
\end{document}